# Wideband Balanced Photodetectors for Classical and Quantum Light Detection from Optical, EUV, to X-rays

IVAN RYGER[1], TERRY BROWN[1], DINA S. EISSA[2], AND CHEN-TING LIAO[2,*]

[1]*JILA, University of Colorado and NIST, 440 UCB, Boulder, Colorado 80309, USA*
[2]*Department of Physics, Indiana University, 727 E. 3rd Street, Bloomington, Indiana 47405, USA*
*Liao3@IU.edu

**Abstract:** The rapid development of coherent short-wavelength light sources in the extreme ultraviolet (EUV) and soft X-ray (SXR) regimes has created a growing need for advanced optoelectronic detection capabilities, particularly for quantum-noise-limited measurements, microelectronics and semiconductor metrology, and emerging quantum information applications. However, extending balanced photodetection to these wavelength regimes is severely hindered by a fundamental bandwidth-noise trade-off imposed by the exceptionally large junction capacitance of EUV-SXR silicon photodiodes. Here, we report a novel wideband photoreceiver architecture that overcomes this bottleneck via a bootstrapped transimpedance amplifier design. By leveraging a low-noise junction field-effect transistor interface, we effectively isolate the photodiode capacitance and suppress the apparent input capacitance seen by the core amplifier. Combined with active compensation of parasitic feedback reactance, this architecture mitigates the conventional trade-off between detector active area and signal bandwidth. Experimentally, we achieved a system-level input-referred noise floor of 13 $\mathrm{fA}/\sqrt{\mathrm{Hz}}$, closely approaching theoretical thermal limits. Furthermore, we achieved a six-fold extension in signal-to-noise limited bandwidth and, through the implementation of a novel grounded field plate, demonstrated a common-mode rejection ratio (CMRR) exceeding 30 dB up to 100 kHz. This highly scalable, silicon-based architecture effectively bridges the short-wavelength detection gap, establishing a robust experimental platform for next-generation quantum-noise-limited and quantum-enhanced X-ray measurement, as well as ultra-sensitive inspection and metrology applications in high-numerical-aperture EUV lithography.

## 1. Introduction: The Detection Gap in the Short-Wavelength Regime

The short-wavelength regime of light, spanning deep ultraviolet to X-rays (wavelength < 200 nm; photon energy > 6 eV), has seen rapid advances in coherent and partially coherent light sources in recent years. Key examples driving this revolution include laser-driven high harmonic generation (HHG) sources enabling femtosecond-to-attosecond spectroscopy [1–3], laser-plasma extreme UV (EUV) sources critical for semiconductor lithography [4], and facility-scale synchrotrons and X-ray free-electron lasers [5,6] These sources have opened new frontiers in probing electron dynamics and nanostructures with unprecedented temporal and spatial resolution [7,8].However, a critical technological gap remains in the corresponding detection infrastructure. Although balanced photodetector designs are routine in the optical domain [9–12], the extreme junction capacitances of EUV-soft X-ray (SXR) diodes have hindered even basic balanced detection at high bandwidths. Balanced photodetection fundamentally outperforms direct signal amplification by canceling common-mode laser noise and isolating weak signals from large DC backgrounds [13–15]. As shown in Fig.1(a), balanced photodetection serves two distinct but equally important roles: direct balanced detection (top) enables ultra-sensitive classical measurements, while balanced homodyne/heterodyne detection (bottom) mixes a faint signal with a bright local oscillator enabling phase-sensitive measurements of optical field quadrature, and, more generally, characterization of quantum states of light. [16–19].

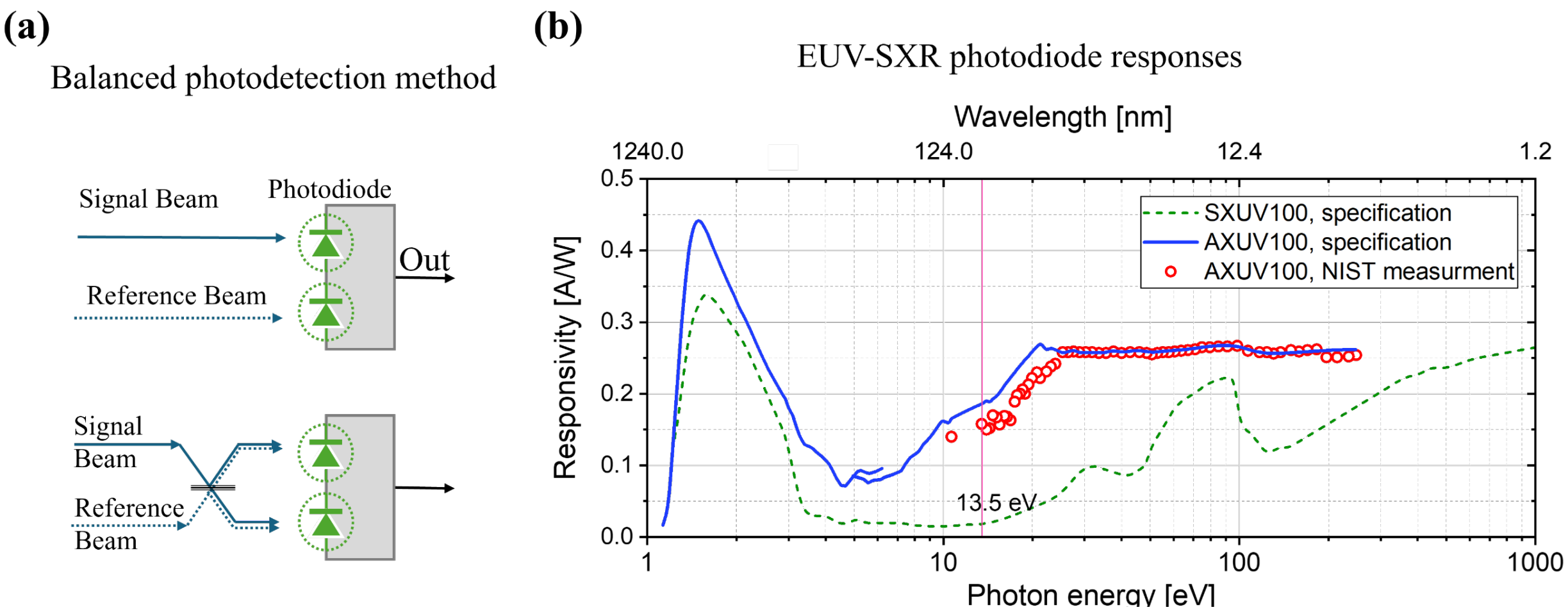

Fig. 1. (a) The basic concept of balanced photodetection in a direct (top) and homodyne (bottom) measurement. (b) Intrinsic spectral responsivity of existing silicon-based photodiodes (AXUV, SXUV) that can detect near-infrared, visible, ultraviolet, EUV, and SXR spectral ranges of light (~1-1,100 nm). Responsivity is measured by photocurrent output in Amperes given optical power input in Watts (A/W) at different wavelengths. Solid blue line: nominal specifications of AXUV100 from the manufacturer; dash green line: nominal specifications of SXUV100 from the manufacturer; red circles: experimentally measured data via NIST provided calibration service on a AXUV100 photodiode.

Bridging this gap is essential for both basic science and industrial application. As effort advance toward shot-noise limited (quantum-noise limited) or sub-shot-noise limited (quantum enhanced) measurements in the X-ray regime [20–22], high-performance balanced detectors are required to resolve minute photocurrent signals, including measuring the anticipated squeezed or entangled short-wavelength EUV and SXR light via balanced homodyne detection [23,24]. Simultaneously, the semiconductor industry's transition to 2-nm and 1-nm nodes via High-NA (numerical aperture) EUV lithography necessitates ultra-sensitive EUV metrology for wafer and photomask inspection [25–27]. In this work, we report on a novel EUV-SXR detector architecture designed to address these specific challenges, offering a pathway to quantum-limited detection in the short-wavelength regime.

## 2. Challenges: The Bandwidth-Noise Trade-off in EUV and X-ray photodetection

One of the primary obstacles to high-performance EUV detection lies not in the availability of short-wavelength-sensitive detector materials, but in the lack of practical wideband, low-noise balanced photodetectors for these wavelengths. Wide-bandgap semiconductors such as silicon carbide, boron/aluminum nitride [28], and diamond [29] are often used for wavelengths below 200 nm because of their superior radiation hardness and low dark current. However, silicon-based photodiodes remain the NIST transfer standards across the 5–254 nm spectral region [30] owing to their near-unity internal quantum efficiency and highly stable, well-characterized spectral responsivity. Owing to these advantages, this work focuses on silicon photodiodes and addresses the challenge of extending them to wideband balanced photodetection. Silicon-based photodetectors such as the AXUV (Absolute Extreme Ultraviolet) and SXUV (Stable Extreme Ultraviolet) photodiodes, originally developed by Gullikson et al. in the 1990s, use shallow delta-doped p-n junctions formed by chemical vapor deposition to suppress carrier recombination near the surface. AXUV employs a nitrided silicon oxide passivation layer, whereas SXUV uses a metal-silicide passivation window (e.g., TiN) [31,32].They also offer a broad dynamic range of measurable power (10 pW–10 mW) and spectral response from hard X-rays (5–10 keV) to infrared (~1.1 eV; 1,100 nm) making them the standard for total flux integration across astronomy [33,34], EUV lithography [35,36] and plasma fusion research [37–40]. AXUV photodiodes are radiation-hardened and engineered for stability in high-flux environments [41], and they can provide nearly 100% internal quantum efficiency in the EUV-SXR spectral ranges [42,43]. The AXUV was commercially co-developed and distributed by International Radiation Detectors (IRD) Inc (now defunct). To maximize light collection efficiency and withstand high flux (rad-hard), AXUV and SXUV photodetectors employ large active areas, often up to 10 mm x 10 mm. Figure 1(b) shows nominal specifications of spectral responsivities of AXUV and SXUV photodiodes. Note that the discontinuity of the blue lines, extracted from manufacturer's specifications, is likely due to measurement or calibration errors. An experimentally measured spectral responsivity of an example AXUV photodiode (NIST-calibrated measurement, AXUV-100G, Opto Diode Corp.) is also shown (red circles). The NIST measurement was based on direct comparison of photocurrents with NIST-calibrated secondary standard photodiode(s) of the same type, all detectors measured under identical illumination. The beam size on the sample photodiode was ~ 2 x 3 mm$^2$ (illumination wavelength 5-50 nm (~24.8-248 eV)) or 3 x 5 mm$^2$ (wavelength 52-254 nm (~4.8-23.8 eV)), centrally positioned at normal incidence. The sample photodiode was operated without bias and positive photocurrent was measured in the anode circuit. NIST calibration method for AXUV is described in Ref. [44].

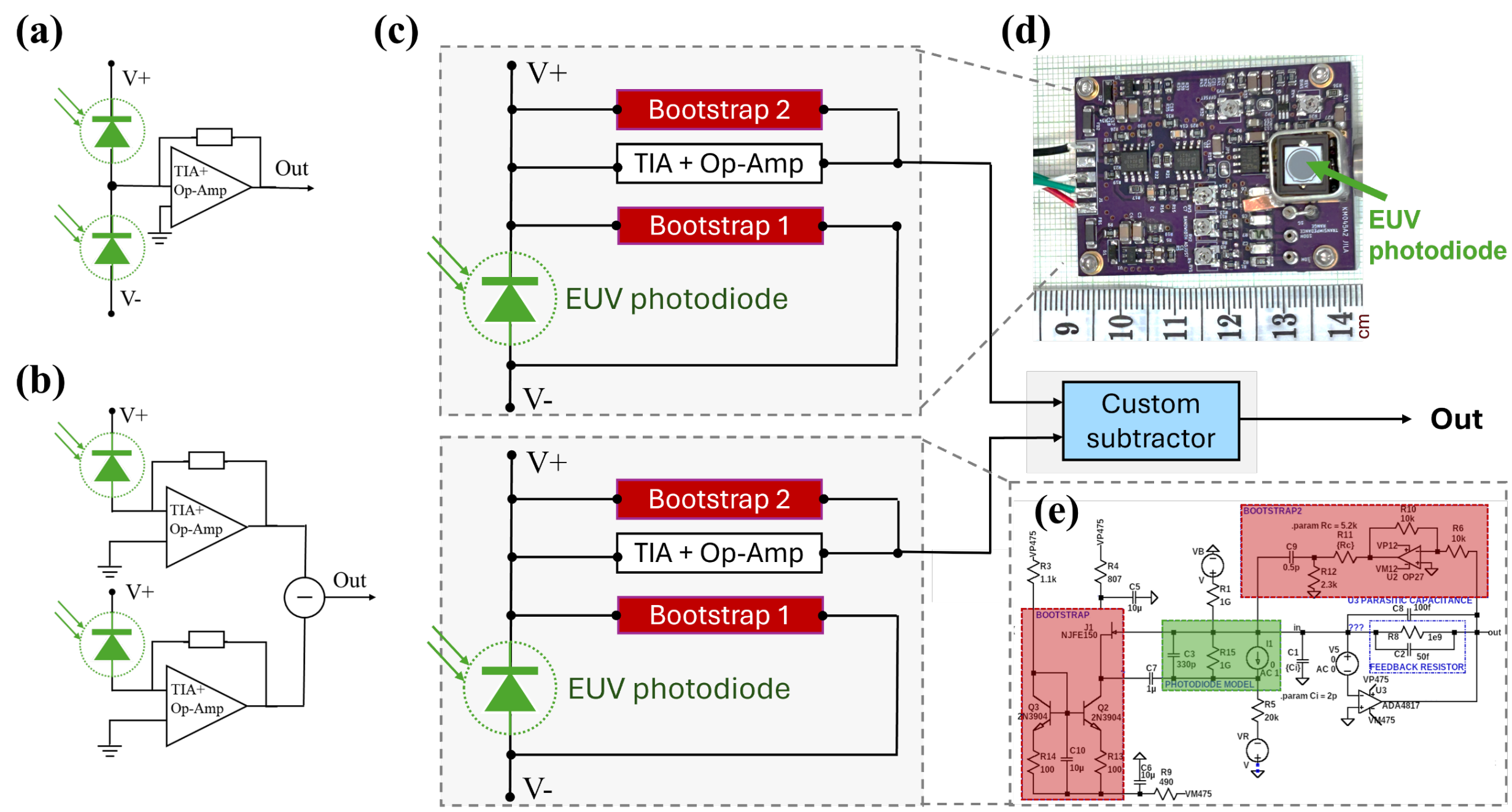

Fig. 2. (a) Balanced photodetection based on current-subtracting-first design. (b) Balanced photodetection based on variable-gain-first design. (c) Our designed photodetector system for optical, EUV, to soft X-ray spectral ranges. (d) Photograph of our fabricated photodetector system (only one photodetector is shown). (e) Simplified schematic circuit diagram of the photodetector.

There are two common, but different types of balanced photodetector design as shown in Figure 2(a) and 2(b), which are current-subtracting-first design and variable-gain-first design, respectively. Given the extremely weak photocurrent signals expected and the bandwidth-noise trade-off, we choose to use the variable-gain-first design (Fig. 2(b)) in our photodetector system. Figure 2(c) illustrates a simplified schematic of our balanced photodetector system design. Figure 2(d) shows a photograph of our final photodetector (only one detector/one channel is shown), where its simplified circuitry is illustrated in Fig.

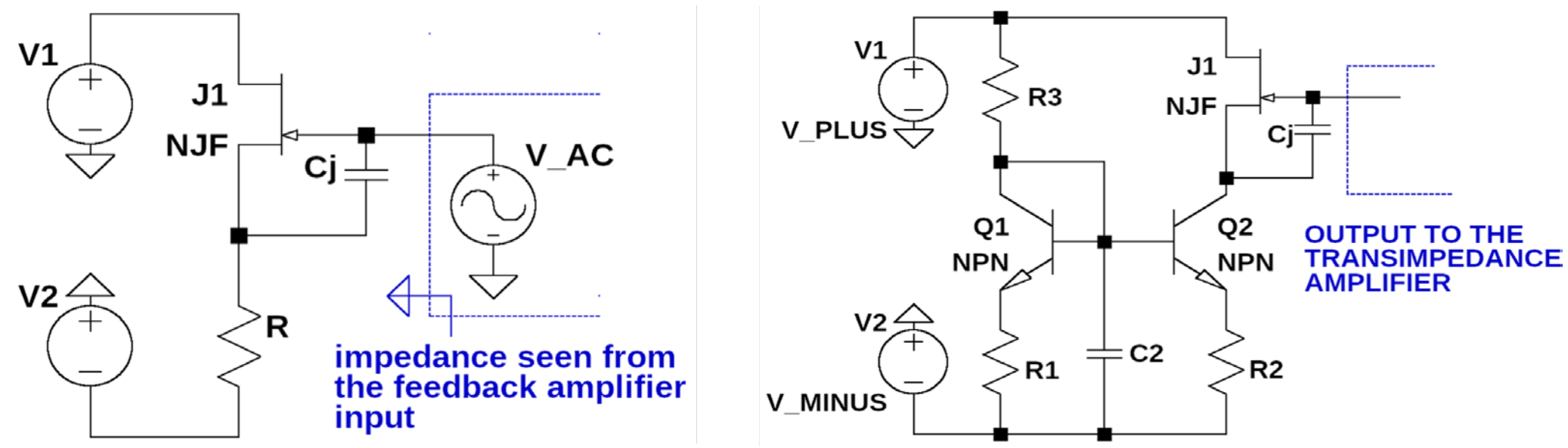

Fig. 3(a). Fundamental bootstrap circuit concept and its impedance seen by the rest of the circuitry. NJF: N-channel JFET. (b) Bootstrap circuit biased by a constant current source formed by a Widlar current mirror. Photodiode junction is represented by capacitance $C_j$. Additional components R1, R2, and C2 suppress the noise contribution of the shot noise originating from the diode junction of Q1 and reduce the voltage gain of Q2, reducing the base resistance noise contribution. NPN: negative-positive-negative transistor.

2(e). The large-area AXUV/SXUV photodiode design creates a fundamental trade-off: although it is beneficial for flux integration, it also gives rise to a large junction capacitance ($C_j$) that severely restricts readout performance. In a standard Transimpedance Amplifier (TIA) topology [45], the signal-to-noise ratio (SNR) is maximized when the feedback resistor's thermal noise is negligible compared to the signal shot noise. This condition yields a lower bound for the optimal feedback resistance: $R_f > 2kT/eI_{ph}$, where $I_{ph}$ is the photocurrent, $k$ is the Boltzmann constant, $T$ is the temperature, and $e$ is the electron charge. For low-light applications, this necessitates a large feedback resistance [46]. However, the interaction between the large photodiode capacitance and the amplifier's input voltage noise ($e_n$) causes "noise peaking”. The noise spectral density transitions from a constant floor to a region proportional to frequency, beginning at a corner frequency determined by the total input capacitance. This pushes the loop gain toward oscillation, typically requiring a compensation capacitor ($C_f$) that further limits the usable bandwidth. However, conventional circuit solutions fail in this low-current regime. A common method to decouple photodiode capacitance is to insert a bipolar junction transistor (BJT) in a common-base configuration. Nevertheless, lowering the dynamic input impedance ($r_e$) of the BJT to a useful level requires a bias current that introduces prohibitive shot noise. For example, for a signal photocurrent at 10 pA (~$10^7$ photons/s in the visible light range). The associated signal shot noise is approximately 1.8 $\mathrm{fA}/\sqrt{\mathrm{Hz}}$. At this current level, $r_e$ is estimated to be ~2.6 GΩ [47]. If photodiode capacitance $C_i$ = 100 pF, the resulting 3 dB bandwidth is a mere 0.6 Hz. To achieve a target measurement bandwidth of 100 kHz, $r_e$ must drop to ~16 kΩ which requires a bias current of 1.6 $\mu A$. Even assuming a transistor with a high current transfer ratio β =100, the base current would be 16 nA, generating a shot noise of 72 $\mathrm{fA}/\sqrt{\mathrm{Hz}}$—approximately 40 times higher than the shot noise of the signal itself. Therefore, a new approach is required to suppress the effect of the capacitance without drowning the faint signal in noise.

The limited choices of EUV-SXR silicon photodiodes and the need to maximize measurement bandwidth motivated us to investigate circuit topologies that minimize the impact of photodiode capacitance on both bandwidth and noise performance. Our goal was to approach photodiode-limited noise up to frequencies of approximately100 kHz. In these devices, charge carriers generated through interband processes, as well as through impurity and trap levels within the space-charge region, are swept by the built-in junction field and contribute to current fluctuations at the electrical terminals. Because the generation-recombination current depends exponentially on temperature, even small variations in ambient conditions can significantly affect the achievable noise floor**.**

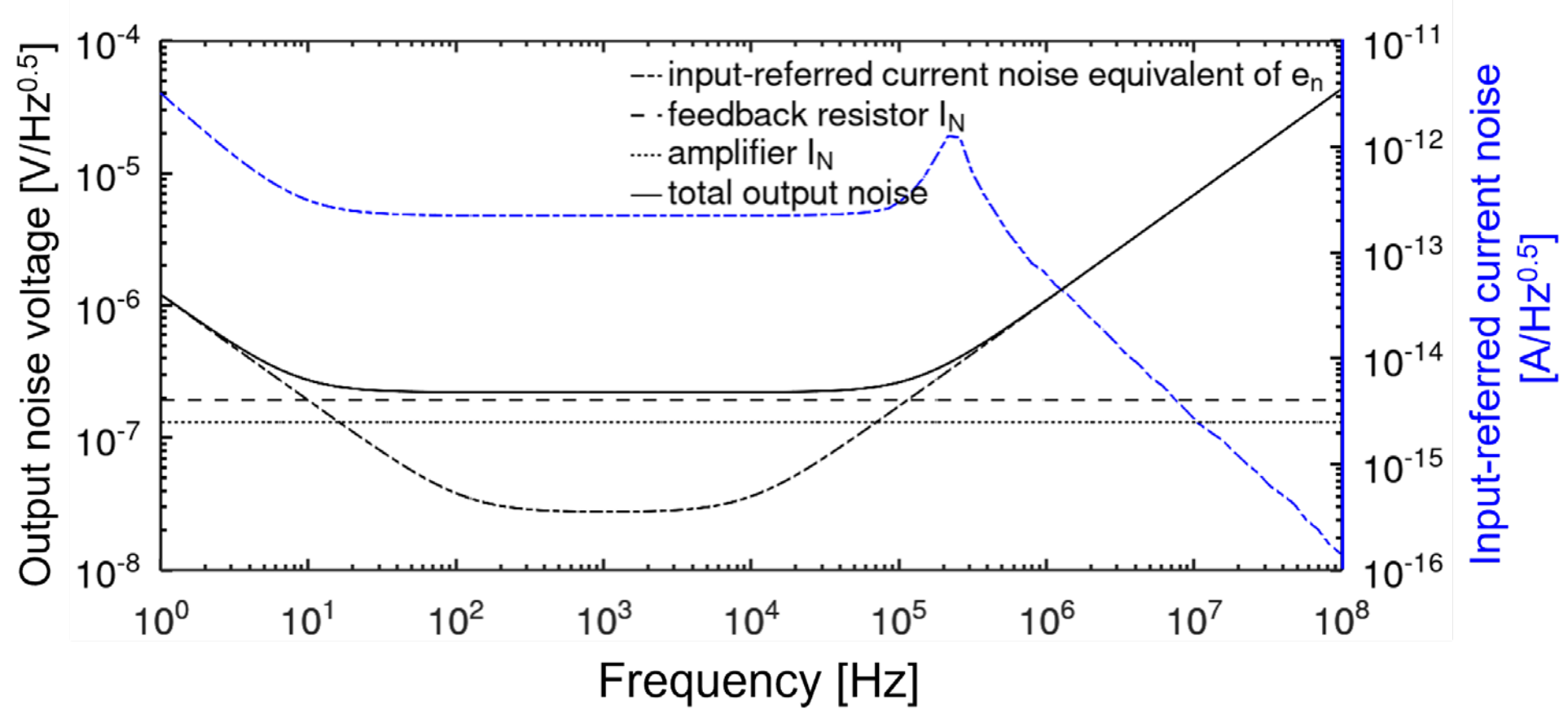

Fig. 4. Calculated minimal attainable noise floor without considering photodiode capacitance. Dash-line: thermal resistor noise limit, dots- amplifier current noise contribution. Blue dash-dot: output-referred voltage noise, dark solid line: calculated current noise spectral density referred to the amplifier input. Dark dash-dot line: contribution of the amplifier voltage noise multiplied by the noise transfer function of the external circuitry.

**The Solution: Bootstrapped Transimpedance Amplifier (TIA) Architecture.** Typical balanced photodetectors (or balanced photoreceivers) either use current-subtracting-first design or variable-gain-first design, as mentioned earlier and shown in Fig. 2(a) and 2(b), respectively. To overcome the fundamental bandwidth-noise trade-off imposed by the large junction capacitance of EUV photodiodes, a standard TIA topology is insufficient. In this section, we detail the implementation of a novel bootstrapped architecture, as shown in our new design in Fig. 2(c), which also illustrates a simplified schematic of our balanced photodetector system design. Figure 2(d) shows a photograph of one photodetector (one channel), and Fig. 2(e) shows its simplified circuit diagram. Our auto-balanced photodetector system has two photodetectors or channels, where the individually TIA signal outputs are then subtracted in a custom subtractor (blue box in Fig. 2), resulting in the final signal readout that exceeding 30 dB common-mode rejection ratio (CMRR) up to 100 kHz signal frequency. Our design strategy prioritizes three active interventions: (1) suppressing the apparent input capacitance via a low-noise junction field-effect transistor (JFET) interface, (2) optimizing the core amplifier for high gain-bandwidth-product operation, and (3) actively compensating for parasitic feedback reactance. Below we discuss these three interventions.

**3.1 The Core Solution: Junction Field-Effect Transistor Bootstrapping.** The primary feature of this design is the isolation of the photodiode capacitance ($C_j$) from the TIA input using a bootstrap buffer. By employing a voltage amplifier with near-unity gain, the AC potential across the photodiode terminals is maintained near zero, effectively reducing the "apparent" capacitance seen by the external circuit. Figure 3(a) shows the fundamental bootstrap circuit concept and its effective input impedance as seen by the rest of the circuit. Designing a bootstrap circuit for low-photocurrent detectors requires careful consideration of intrinsic noise. The primary criterion is that the bias current of the bootstrap transistor must be orders of magnitude lower than the photocurrent to minimize thermal drift and shot noise. Consequently, bipolar transistors are unsuitable for low-current applications; instead, a JFET in a source-follower configuration is typically used. Figure 3(b) shows our analysis of the bootstrap circuit, which yields the input admittance $Y_i = j\omega\, C_j/(1 + R(g_m + j\omega C_j))$. Here, $C_j$ is the photodiode capacitance, $R$ is the bootstrap resistor, $g_m$ is the transconductance of the JFET, and ω=2πf is the angular frequency. This relationship shows that for frequencies below $g_m/(2\pi C_j)$, capacitance reduction is roughly proportional to the factor $R_{gm}$, where $g_m$ is the transconductance.

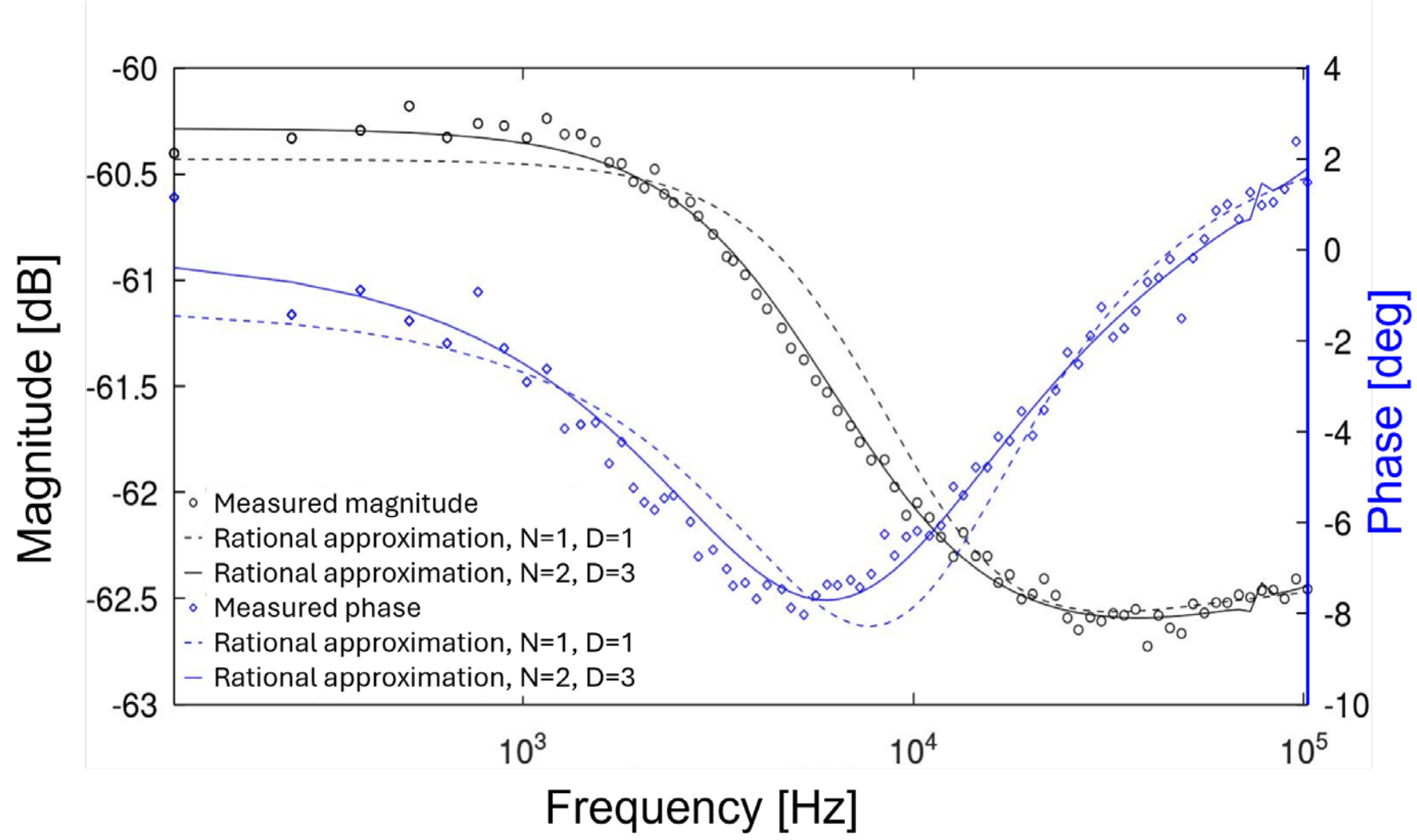

Fig. 5. Experimental measured voltage transfer function (TF) magnitude (black; left vertical axis) and phase (blue; right vertical axis) for the high-GΩ value feedback resistor. Solid symbols are measured data points; lines fitting by rational polynomial approximation, numerator (N; N = 1 or 2) and denominator (D; D = 1 or 3) degrees.

**Active Device Selection.** The choice of the bootstrap transistor is critical; its bias current must be orders of magnitude lower than the signal photocurrent to minimize thermal drift and shot noise. We selected the JFE150 (Texas Instruments), an ultra-low-noise n-channel JFET. This component features a high transconductance ($g_m$ = 68 mS at $V_{gs}$ = 0 V) and ultra-low leakage current ($I_b$ < 10 pA), ensuring that the bootstrap circuit itself does not introduce significant noise penalties.

**Active Load Implementation**. To achieve the high load resistance required for near-unity gain without resorting to excessively high supply voltage rails, we implemented an active load using a Widlar current mirror. This constant current source offers superior stability against temperature variations compared to passive resistor biasing. Emitter degeneration was employed to reduce noise amplification from the mirror transistors.

**Low-noise current source.** The current source itself introduces noise. Fluctuations in the drain current cause variations in $V_{gs}$, inducing a displacement current through the photodiode junction capacitance $C_j$. This noise source scales with frequency and appears at the TIA input with a power spectral density of $S_{id}(\omega) = (\omega\, C_j/g_m)^2 S_i(\omega)$, where the $S_{id}(\omega)$ is the current noise density of the current source and $S_i(\omega)$ represents the current noise density of the displacement current flowing through the attached $C_j$. To mitigate the Johnson-Nyquist noise from the base resistance and shot noise from the mirror transistors [47,48], we employed emitter degeneration (adding series resistance) to reduce transconductance and noise amplification [49]. Furthermore, an RC filter (R1-C2 in Fig. 3(b)) was added to reduce output noise, bringing it close to the theoretical shot-noise limit of a single collector.

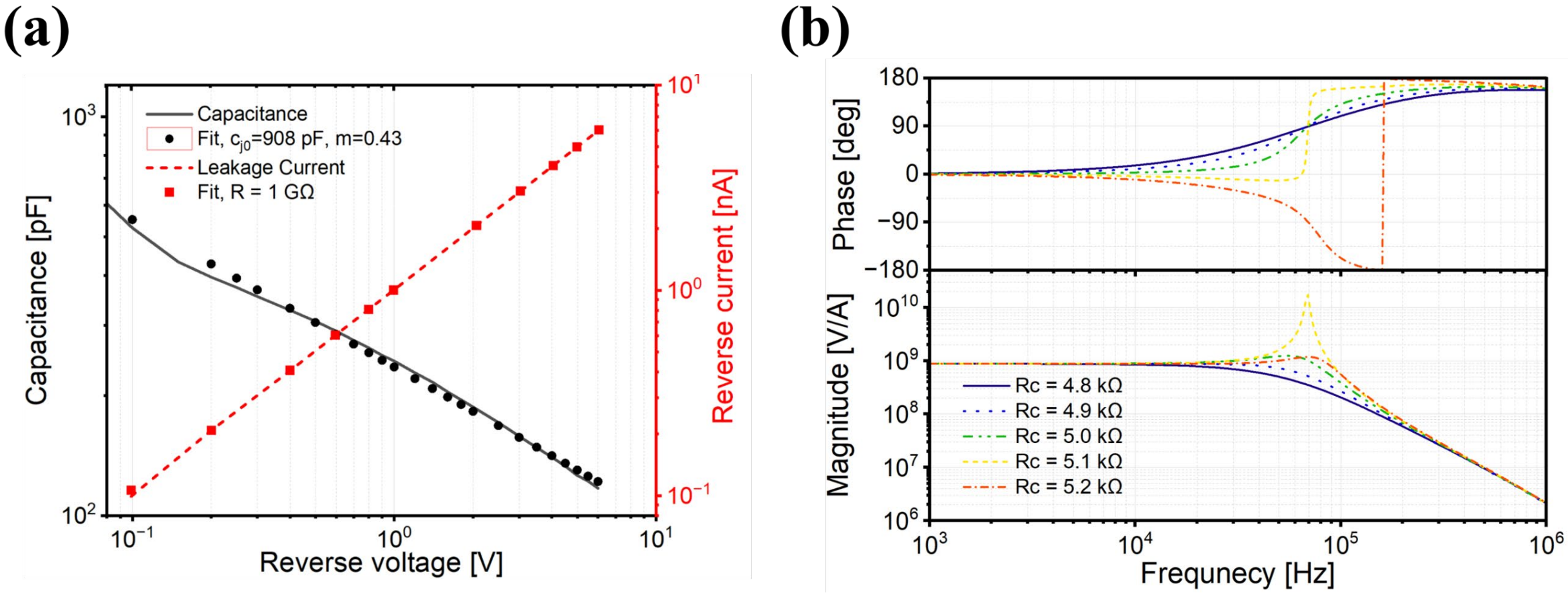

Fig. 6. (a) Experimentally measured photodiode junction capacitance (black circles) and reverse leakage current (red squares) alongside with fitting curve (red dash line). (b) Frequency dependence of magnitude (bottom) and phase (top) of transimpedance at difference varying bootstrap feedback ratio, given by resistance $R_C$.

**3.2 Operational Amplifier Selection and Noise Limits.** With the input capacitance effectively suppressed, the noise floor is determined by the core TIA. We employed a semi-analytic modeling approach to select the optimal operational amplifier (op-amp), extracting gain and phase characteristics directly from manufacturer data.

Based on this analysis, the ADA4817 (Analog Devices) was selected for its optimal combination of low bias current noise, low input capacitance, and high gain-bandwidth product. Our theoretical modeling (Fig. 4) indicates that with a feedback resistor of $R_f$ = 1 GΩ, the estimated equivalent input-referred current noise is 7.5 fA/$\sqrt{\text{Hz}}$ at 100 kHz, rising to ~75 fA/$\sqrt{\text{Hz}}$ at 1 MHz. This calculation assumes a total input capacitance of 3 pF and establishes the theoretical lower bound for the system.

**3.3 Secondary Optimization: Feedback-Parasitic Compensation.** While bootstrapping suppresses the contribution of the input capacitance, the overall bandwidth remains limited by the parasitic capacitance ($C_f$) across the large-value feedback resistor. Using a dual-channel fast-Fourier transform network analyzer (SR780, Stanford Research Systems), we characterized this parasitic capacitance by configuring a voltage divider consisting of the high-impedance surface-mount thick-film resistor under test and the known input impedance of the measurement instrument (1MΩ, 50 pF). The complex frequency transfer function was measured with an internal pseudo-random noise source providing excitation. To eliminate the contribution of stray fringing fields from the connection leads, both input and output leads were shielded to ground potential using copper tape. The transfer function was recorded over a frequency range of 100 Hz to 100 kHz. Subsequently, the measured data points were approximated using a rational polynomial fit of the form: $T(s) = (b_0 + b_1 s + b_2 s^2)/(a_0 + a_1 s + a_2 s^2 + a_3 s^3)$, where $s = j\omega$ and $\omega = 2\pi f$. In the reduced-order case where coefficients $a_2$, $a_3$, and $b_2$ are negligible, the rational approximation describes the frequency response of a simple two-impedance divider. Under these conditions, the coefficient ratios correspond to physical circuit parameters: $b_0/a_0 = R_i/R_f = k$ represents the DC voltage transfer; $b_1/b_0 = R_f C_f = \tau_f$ represents the time constant of the feedback resistor; $a_1/a_0 = R_i(C_i + C_f) = \tau_i$ represents the time constant dominated by the instrument input impedance. This simplified transfer function yields only approximate values for the parasitic reactance, as the physical device tends to exhibit distributed reactance effects similar to a transmission line, causing deviations from the lumped-element model (Fig. 5, symbols vs. dashed line). Notably, the deviation between measured points and the simple pole-zero approximation is most significant in the 1 kHz to 10 kHz region, decreases at higher frequencies. A third-order polynomial fit follows the data trend more closely, suggesting that a higher-order cascaded R-C compensation topology would improve the flatness of the transimpedance frequency response. Comparing measurement techniques, earlier characterization attempts using an oscilloscope for magnitude and phase estimation yielded an approximate feedback capacitance of $C_f \sim 0.045$ pF. However, the network analyzer measurement, utilizing denser data points and a robust polynomial fit, resulted in a refined estimate of $C_f \sim 0.027$ pF.

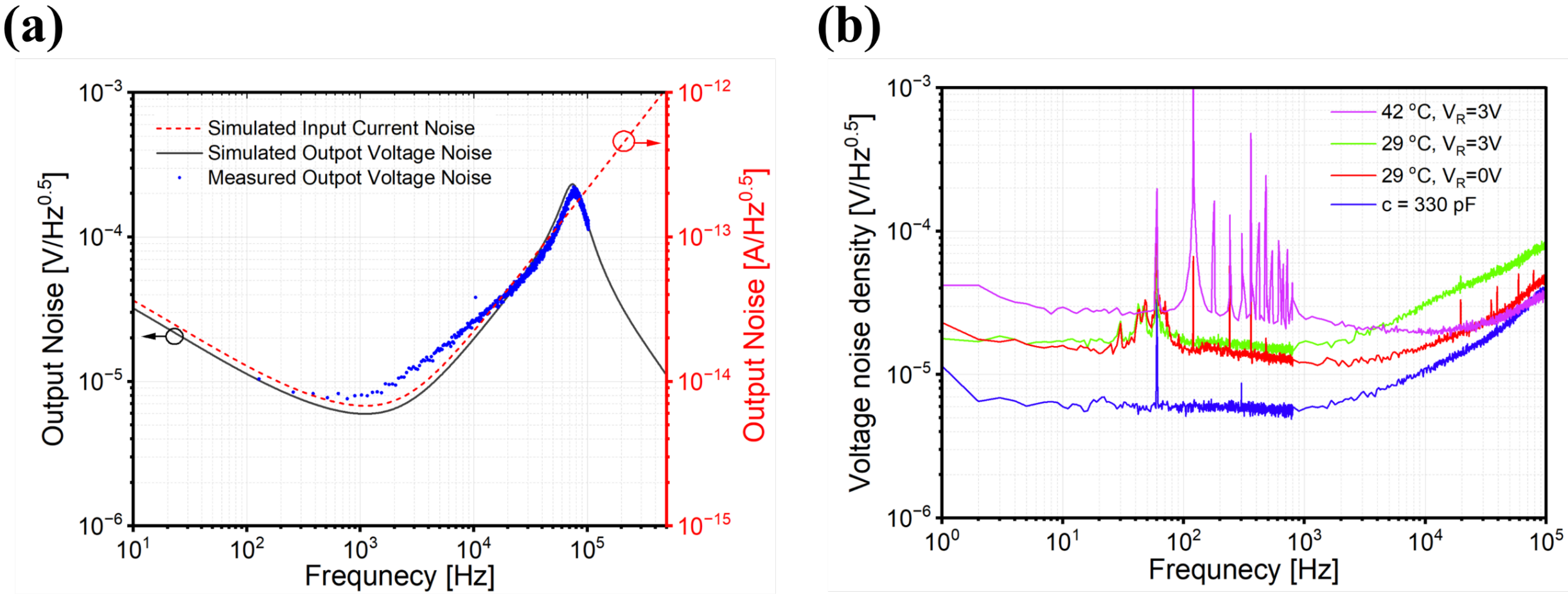

Fig. 7. (a) Comparison of experimental noise floor measurement with numerical SPICE simulation. The solid and broken lines show the simulated output voltage noise (black solid line) at the photodetector amplifier output and input-referred noise current noise (red dashed line), respectively. Experimental measured data points are represented by blue points. (b) Comparative measurements of photodetector noise frequency dependence. Bottom trace in blue: photodiode replaced by 330 pF capacitor. Upper traces: photodiode with applied reverse bias voltages and different temperatures.

## 4. Experimental Validation and Results

To experimentally validate the proposed architecture, we employed a hierarchical characterization protocol. We first isolated the intrinsic properties of the EUV photodiode to quantify the capacitive load. Second, we verified the electronic noise floor against theoretical limits using dummy loads. Finally, the complete system was integrated into a balanced optical homodyne setup to evaluate signal fidelity and common-mode rejection.

**4.1 Component Characterization:** Prior to circuit and system integration, the AXUV20HS photodiode, as shown in the photograph of Fig. 2(d), with intrinsic response time of 2 ns (Opto Diode Corp.) was characterized to determine its junction capacitance ($C_j$) and dark current profile. Accurate knowledge of $C_j$ is critical, as it dictates the efficacy of the bootstrapping network. Measurements were performed using a Stanford Research SR720 LCR meter for capacitance profiling and a Keithley 615 electrometer for high-precision leakage current measurements. As illustrated in Fig. 6(a), the unbiased photodiode presents a massive junction capacitance due to its large active area (100 mm$^2$). However, applying a reverse bias of approximately -3.5 V reduced the junction capacitance roughly six-fold compared to the unbiased state. The leakage current data ($I_{dark}$) and capacitance were fitted to a theoretical model. The capacitance followed a modified power law $C_j(V) = C_{j0}(1 + V/V_j)^{-m}$. Note that this modified equation is phenomenological rather than physically derived, it is sufficient for modeling purposes. This behavior is consistent with the depletion region expansion in the shallow p-n junctions characteristic of SXUV/AXUV photodiode. The results confirm that while the large area is necessary for flux integration, moderate reverse biasing is a prerequisite for achieving sufficient bandwidth, serving as the first stage of capacitance management before electronic bootstrapping is applied. Figure 6(b) shows the effect of secondary bootstrap circuit, compensating the parasitic capacitance of the feedback resistor. By changing the resistance $R_C$ the feedback transfer ratio at high frequencies reduces the phase lag (caused by the parasitic feedback capacitance), leading from over-dampened case in one extreme to under-dampened oscillatory behavior. By optimal setting of the value $R_C$, the feedback bandwidth is improved.

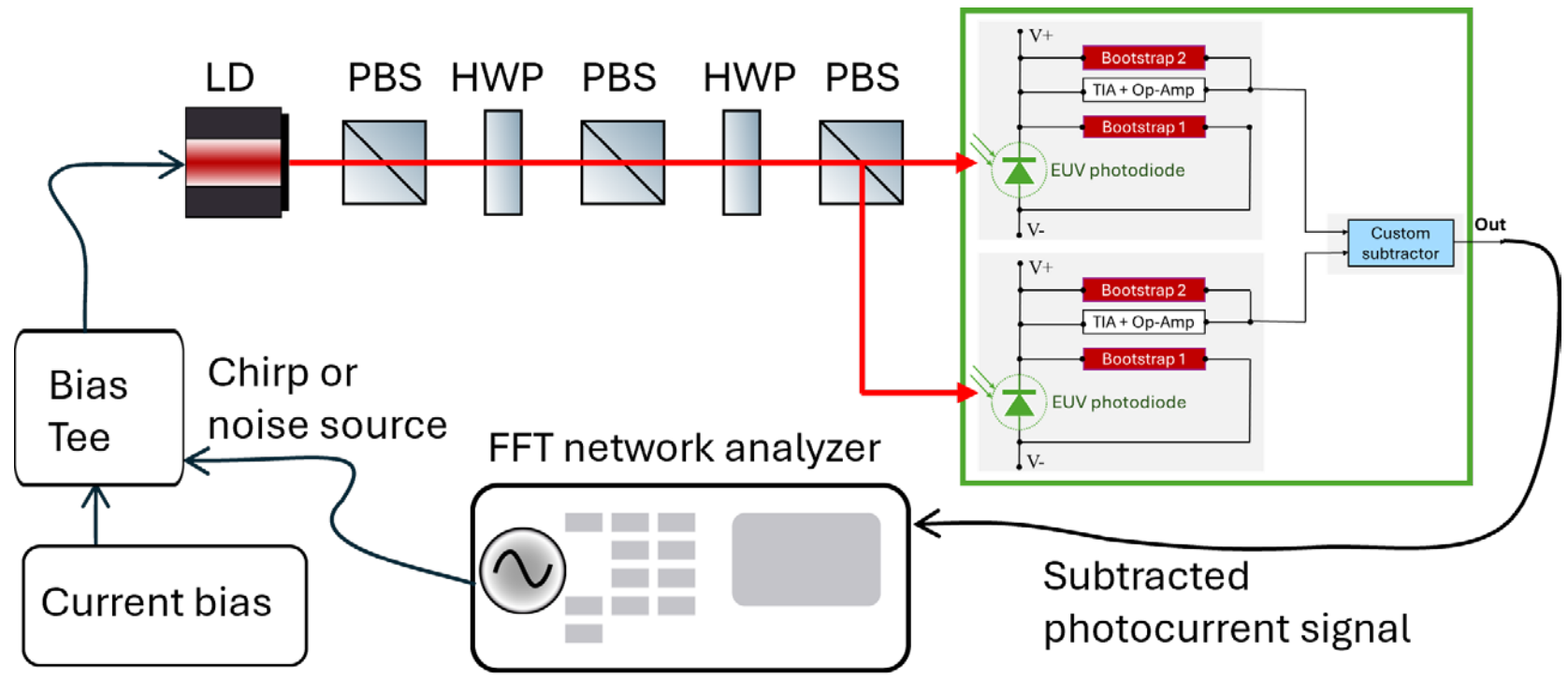

Fig. 8. Schematic diagram of balanced photodetection testing setup. The green block indicates our designed and fabricated balanced photodetector system as shown in Fig. 2(c) LD: laser diode, PBS: polarizing beam splitter, HWP: half-wave plate.

**4.2 Circuit Noise Performance.** The noise performance of the transimpedance amplifier was evaluated in a shielded environment to determine the input-referred noise current density.

**Baseline Verification** (Dummy Load): To rigorously isolate the amplifier's intrinsic noise from the photodiode's shot noise, the photodiode was initially replaced by a fixed, low-loss capacitor simulating the depleted junction capacitance. At low frequencies, the measured input-referred noise current was 7 $\mathrm{fA}/\sqrt{\mathrm{Hz}}$. This aligns within measurement uncertainty to the theoretical limit of 6 $\mathrm{fA}/\sqrt{\mathrm{Hz}}$, which is derived from the combined shot noise of the JFET gate leakage and the op-amp bias currents. This confirms that the addition of the JFET bootstrapping network adds negligible thermal penalties to the system.

**System Noise** (Photodiode Connected): Upon connecting the AXUV20HS photodiode, the broadband "white" noise floor increased to ~13 $\mathrm{fA}/\sqrt{\mathrm{Hz}}$, as shown in Fig. 7. This increase is attributed primarily to the shot noise of the dark current, with smaller contributions from the thermal noise of the photodiode series resistance. Figure 7(a) shows our experimental measured output noise floor (blue dots) from 10Hz to 50kHz and its comparison to numerical SPICE simulation (red and black lines). The black-solid and red-broken lines show the simulated output voltage noise at the photodetector amplifier output and input-referred noise current noise, respectively. Figure 7(b) shows an experimental measured photodetector noise frequency dependence. The bottom trace in blue shows photodiode replaced by 330 pF capacitor, and the upper traces are photodiode with applied reverse bias voltages at different temperatures.

**Thermal Limits.** Given the high sensitivity of the JFET and the large dark current of EUV photodiodes, thermal stability is a primary concern. We subjected the prototype to controlled thermal stress testing. At an elevated temperature of 42 °C as shown in Fig. 7(b), the photodiode leakage current rose to approximately 2 nA. This increase in shot noise raised the broadband noise floor to 28 $\mathrm{fA}/\sqrt{\mathrm{Hz}}$, degrading the SNR. More critically, exposure to laboratory air at 50 °C resulted in an irreversible increase in leakage current, indicating permanent degradation of the shallow p-n junction interface. These results confirm that while the electronic architecture is robust, practical deployment in synchrotron or lithography environments will require active thermal management (e.g., Peltier cooling) and hermetic encapsulation to preserve the detector's quantum-limited sensitivity over long operational periods, to mitigate the exponential rise of dark current in silicon detectors.

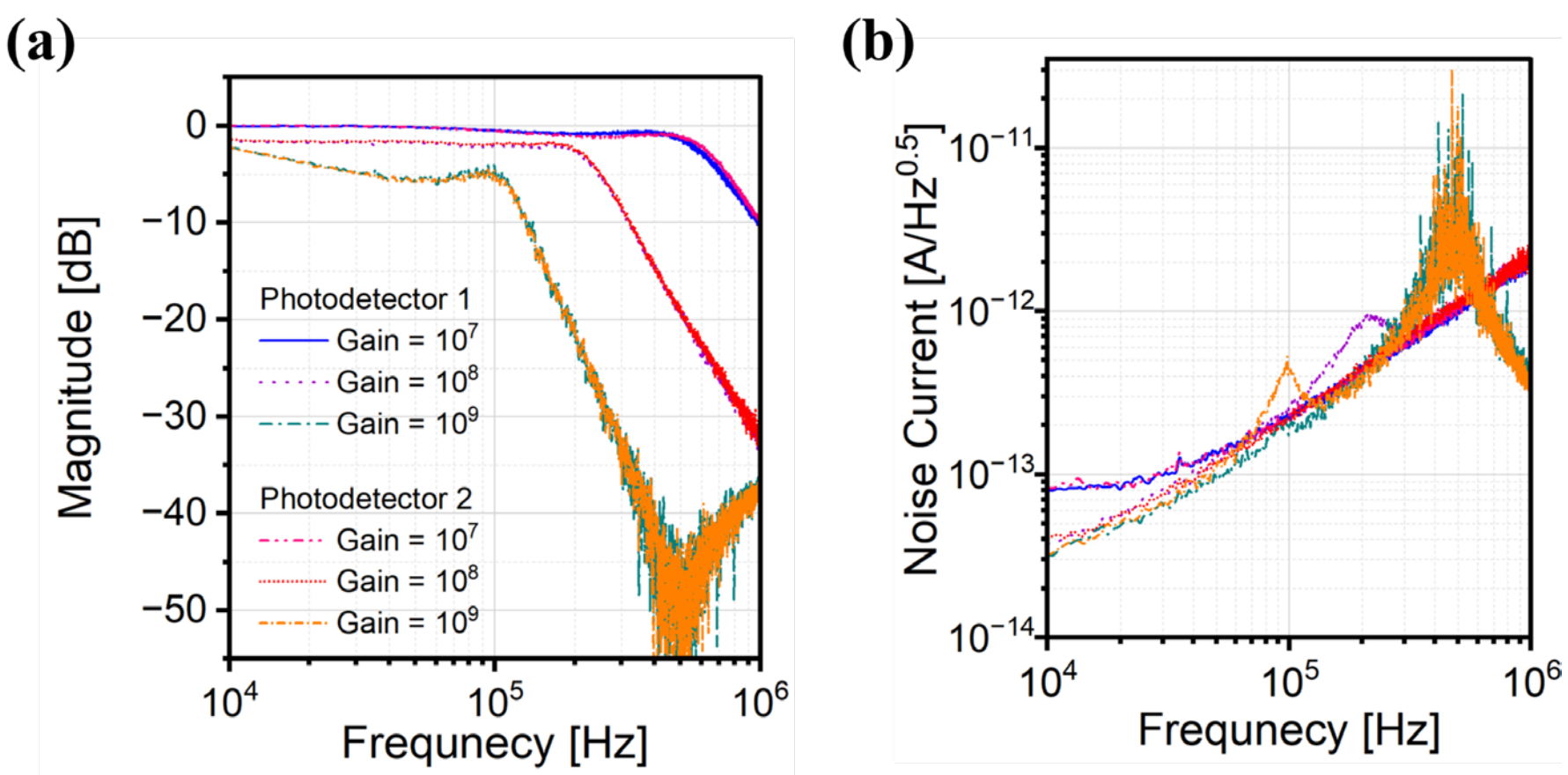

Fig. 9. (a) Frequency response of the transimpedance amplifier for different transimpedance settings (Gain, $10^7$, $10^8$, $10^9$) at both photodetectors. (b) Their associated input-referred current noise spectral density.

**Bandwidth Extension:** The efficacy of the design was most evident in the bandwidth measurements. Consistent with the capacitance reduction observed in Section 4.1, applying a reverse bias of -3 V to the photodiode increased the signal-to-noise-limited bandwidth by a factor of six, confirming the system-level benefit of reduced junction capacitance. Furthermore, by adjusting the secondary bootstrap compensation (Fig. 6(b)), we were able to tune the step response from an overdamped to an underdamped regime, verifying that the circuit can effectively nullify the parasitic feedback capacitance.

**4.3 System Level Tests of Balanced Homodyne Detection:** The photodetector's performance and balancing capabilities were experimentally evaluated in a realistic optical setup. We constructed a balanced homodyne detector testbed using a 635 nm wavelength laser diode as shown in Fig. 8. At 635nm (1.95 eV), the photodiode's intrinsic responsivity is expected to be at ~0.33 A/W, which is comparable to those in the EUV-SXR ranges (~0.25 A/W from 20eV-250eV). In our testing setup, the red laser beam is first sent through a polarizing beam splitter (PBS) to make it linearly polarized. A subsequent half-wave plate (HWP) and another PBS are used to tune the laser power without altering the laser's noise characteristics. This is important because the noise characteristics of a diode laser, which is driven by an electric current source, the current bias block shown in Fig. 8, are laser power dependent—given that the laser power depends on the driving current bias. After that, the laser power is further controlled by another HWP and PBS to split the beam power equally between two photodetector channels. The balanced photodetector system is shown in the green block, which is the same as in Fig. 2(c). The subtracted photocurrent output signal is then acquired by an FFT (Fast Fourier Transform) network analyzer (Agilent/HP 89410A Vector Signal Analyzer) to evaluate the photodetector system's performance. The FFT network analyzer is also used to generate a chirp or noise source, which is mixed via a bias tee with a current bias that drives the diode laser.

**Individual photodetector frequency responses.** The individual photodetector (two channels, photodetector 1 and photodetector 2) frequency responses were measured for three different transimpedance settings, at gain, $10^7$, $10^8$, and $10^9$ V/A, as shown in Fig. 9. The frequency response was approximated again by a ratio of two complex polynomials. This smooth fitting function from Fig. 9(a) was then used in the denominator to calculate the input-referred current noise floor for the transimpedance amplifier. This can be then converted and scaled to the input equivalent noise power through the particular diode responsivity at the given wavelength as shown in Fig. 9(b). As we can observe that the noise current changes from ~30 fA/ √ Hz to ~10 pA/ √ Hz in the frequency range of 10 kHz to 1,000 kHz. The noise current is around 200 fA/ √ Hz at 100kHz.

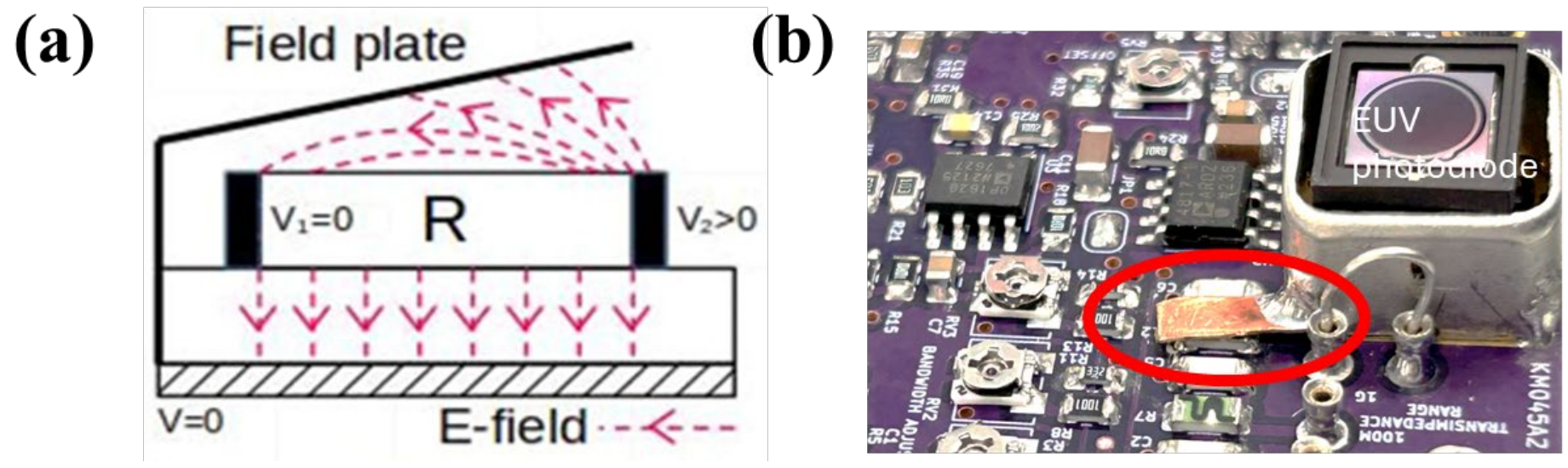

Fig. 10. Schematic showing electric field distribution for a resistor with a field plate. (b) Photograph of how we physical realizes a field plate in a photodetector (red circle).

**Low-noise voltage regulation and thermal management.** Given the sensitivity of low-photocurrent TIAs, stable low-noise voltage regulation is essential to reject power supply noise, which is limited mainly by the photodiode bootstrap circuitry, and by the op-amp's finite power supply rejection ratio to lesser extent. We utilized monolithic low-noise regulators (ADP7118 and ASP7182, Analog Devices). For ultra-low-noise requirements, a standard regulator (e.g., LM317) can be combined with a noise-suppressing bridge circuit. Thermal management is also critical. Linear regulators should operate with minimal voltage drop to reduce dissipation, and low-power op-amps should be used for non-critical stages (buffers, inverters). For further dark current reduction, the photodetector can be mounted on a thermoelectric (Peltier) cooler. Finally, to prevent drift from induced charges and interference from power-line magnetic fields, the photodetector must be enclosed in a Faraday cage connected to the common terminal or the experiment chamber.

**Field plate tuning for two-channel frequency matching.** A critical challenge in high-gain ($10^9$ V/A) balanced detection is to match the parasitic capacitances of the two channels. To address this, we implemented a grounded field plate positioned near the feedback resistor as discussed in Ref. [50]. As shown in Fig. 10, this plate acts as a capacitive shield, diverting a portion of the displacement current from the output to the ground, rather than allowing it to couple back into the high-impedance input node (Fig. 10(a)). By physically adjusting the position of the field plate (Fig. 10(b), we can tune the ratio between longitudinal and transverse fringing capacitance. This mechanical "trimming" allowed for precise matching of the frequency response between the two independent photodiode signal channels.

**Final balanced photodetector system performance results.** With the channel responses matched via the field plate, we can achieve the algebraic difference between their DC-normalized responses to determine the Common Mode Rejection Ratio (CMRR) of the balanced photodetector system. In our experiment, the system demonstrated a CMRR better than 30 dB up to 100 kHz in our optical test, as shown in Fig. 11. Normalized photodetector frequency response from individual photodetectors (photodetector 1, PD1, red dash line, and photodetector 2, PD2, blue dotted line) are measured, as well as the final attainable common mode signal suppression (PD1-PD2 response; green solid line). We noted an apparent decrease in suppression between 100 Hz and 1 kHz; however, this was identified as a measurement artifact arising from the dynamic range limits of the FFT analyzer used for the test, rather than a failure of the detector itself. At other frequency ranges, for example, around 10 kHz, we also achieved approximately 40 dB CMRR.

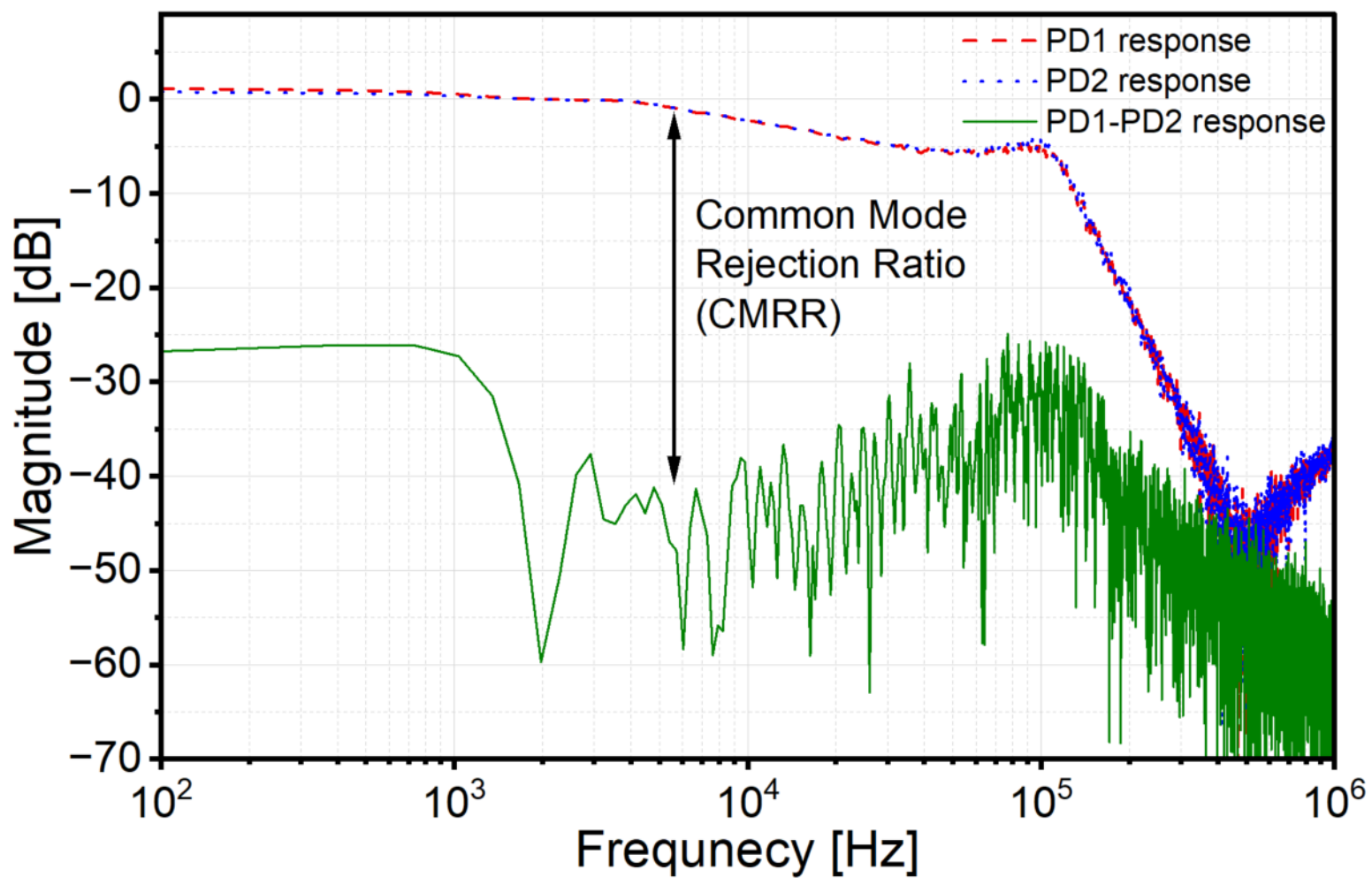

Fig. 11. The final CMRR output performance result. Normalized photodetector frequency response for two photodetectors, PD1 (red dash line) and PD2 (blue dotted line) and attainable common mode signal suppression (PD1-PD2 response; green solid line).

## 5. Conclusion

In this work, we have addressed the critical optoelectronic gap hindering the advancement of short-wavelength quantum sensing and metrology. While coherent sources in the EUV and soft X-ray regimes have matured significantly, the corresponding detection infrastructure has lagged, primarily due to the fundamental physical limitations of available photodetectors. We identified that the large junction capacitance inherent to high-flux EUV-SXR photodiodes acts as a primary bottleneck, severely restricting the bandwidth-noise performance of standard transimpedance topologies. To overcome this, we designed and validated a novel wideband photoreceiver architecture centered on JFET-based input bootstrapping. By isolating the photodiode capacitance from the amplifier input using a low-noise JFET buffer, we successfully suppressed the apparent capacitance seen by the circuit. This architectural innovation, combined with active parasitic feedback compensation, allowed us to break the traditional trade-off between active area and signal bandwidth.

Our experimental results confirm the robustness of this approach across three critical dimensions: (1) In terms of noise performance, we achieved a system-level input-referred noise floor of ~13 $\mathrm{fA}/\sqrt{\mathrm{Hz}}$, closely approaching the theoretical thermal limits. This represents a significant milestone for detecting faint photon fluxes in the presence of massive detector capacitance. (2) In terms of bandwidth extension, the combination of reverse-biasing and electronic bootstrapping resulted in a six-fold extension of the signal-to-noise limited bandwidth, enabling time-domain resolution previously unattainable with large-area sensors. (3) In terms of balanced detection, through the implementation of a novel grounded field plate for fringing capacitance tuning, we demonstrated a Common-mode rejection ratio (CMRR) exceeding 30 dB up to 100 kHz. This high degree of balancing is essential for isolating quantum fluctuations from the noisy backgrounds typical of synchrotron and plasma-based sources. Given the well-established responsivity of these photodiodes from the near-infrared to the soft X-ray regime, we expect comparable detector performance across the EUV and SXR ranges, provided that the readout conditions are otherwise similar. For example, for photon energy at EUV-SXR of 20-250 eV, the spectral responsivity is ~0.25 A/W, which is the same as that in the visible light of ~500 nm.

As indicated by our thermal stress tests, future work will need to incorporate active thermal management strategies, such as Peltier or water cooling, to mitigate excessive thermal noise. In addition, hermetic encapsulation may be required to preserve the detector's quantum-limited sensitivity over extended operational periods and during routine daily operation.

Looking forward, the technology we developed provides the necessary instrumentation foundation for next-generation optoelectronic applications, including X-ray spectroscopy and diffractive imaging [51] and actinic inspection and metrology for High-NA EUV based metrology for lithography nodes [26,52]. Ultimately, this work establishes a scalable pathway toward quantum-limited photodetection spanning the optical to soft X-ray spectrum.

----

**Funding.** This work is supported by the U.S. Air Force Office of Scientific Research (AFOSR), Young Investigator Program (YIP), Award no. FA9550-23-1-0234. Any opinions, findings, and conclusions or recommendations expressed in this material are those of the author(s) and do not necessarily reflect the views of the U.S. Air Force.

**Acknowledgment.** C.-T. L. acknowledges administrative support from Prof. Shu-Wei Huang of University of Colorado Boulder for the YIP grant management.

**Disclosures.** Some co-authors have a relevant provisional patent application related to this work.

**Data availability.** Data underlying the results presented in this paper are not publicly available at this time but may be obtained from the authors upon reasonable request.